\documentclass[twocolumn,showpacs,preprintnumbers,amsmath,amssymb,superscriptaddress]{revtex4}
\usepackage{graphicx}% Include figure files \usepackage[dvips]{color}
\usepackage{dcolumn}% Align table columns on decimal point
\usepackage{bm}% bold math
\usepackage[nooneline]{subfigure}
\usepackage{verbatim}
\begin{document}

\preprint{APS/123-QED}

\title{Influence of disorder strength on phase field models of interfacial growth}

\author{T. Laurila}%
\affiliation{Department of Engineering Physics, P.O. Box 1100, Helsinki
University of Technology, FI-02015 TKK, Espoo, Finland}
%\email{}

\author{M. Pradas}%
\affiliation{Departament d'Estructura i Constituents de la
Mat\`eria,\\ Universitat de Barcelona, Avinguda Diagonal 647,
E-08028 Barcelona, Spain}

\author{A. Hern\'andez-Machado}%
\affiliation{Departament d'Estructura i Constituents de la
Mat\`eria,\\ Universitat de Barcelona, Avinguda Diagonal 647,
E-08028 Barcelona, Spain}

\author{T. Ala-Nissila}%
\affiliation{Department of Engineering Physics, P.O. Box 1100, Helsinki
University of Technology, FI-02015 TKK, Espoo, Finland}
\affiliation{Department of Physics, Brown University, Providence RI 02912-1854, U.S.A.}

 %\email{Second.Author@institution.edu}

\date{May 30, 2008}

\begin{abstract}
We study the influence of disorder strength on the interface roughening process in a phase-field
model with locally conserved dynamics. We consider two cases where the mobility
coefficient multiplying the locally conserved current is either
constant throughout the system (the two-sided model)
or becomes zero in the phase into which the interface advances (one-sided model).
In the limit of weak disorder, both models are completely equivalent and can reproduce the
physical process of a fluid diffusively invading a porous media, where
super-rough scaling of the interface fluctuations occurs. On the other hand,
increasing disorder causes the scaling properties to change to intrinsic anomalous scaling.
In the limit of strong disorder this behavior prevails for
the one-sided model, whereas for the two-sided case, nucleation of domains in front
of the invading front are observed.
\end{abstract}
%
%25.2.
\pacs{Valid PACS appear here}% PACS, the Physics and Astronomy %
                             %Classification Scheme.
%\keywords{Suggested keywords}%Use showkeys class option if keyword
                              %display desired

\maketitle

\section{Introduction}\label{SecI}
Interface growth in disordered systems has been a subject of great interest in non-equilibrium statistical
physics for more than a decade. Many different examples of interfaces growing in
heterogeneous systems have been found in phase separation by nucleation and growth~\cite{cahn65,GU83},
solidification~\cite{langer80,boettinger02}, and fluid invasion in porous media~\cite{AL04,AN04}, among others.

Phase field models of increasing complexity have in recent years been
extensively used in studying interface roughening as well as microstructure formation
\cite{AL04,HH77,elder94,EL01,chen02,LU05,LA06,rost07}. A particularly interesting problem of interface roughening is that
associated with a fluid invasion front moving into a disorder medium~\cite{AL04}, which can be experimentally
studied with the Hele-Shaw cell set-up \cite{SO07,SO05,SO02,SO02b,GE02}. In modeling such a fluid invasion
experiment, two different ways to consider the mobility parameter in a Model B type of phase field model,
called the one-sided and symmetric models, were used by Hern\'andez-Machado \emph{et al.} \cite{HM01}
and Dub\'e \emph{et al.} \cite{DU99,DU00}, respectively. The difference
between these two cases is that in the two-sided model the mobility is constant throughout the system, while in the one-sided
model it becomes zero in the phase which is being invaded.

In this paper our aim is to carry out a detailed analysis about the influence of the strength of the disorder in the two types of models described above.
%in the presence of disordered.
Both cases can be described by a generalized Cahn-Hilliard equation or Model B with
quenched disorder in the background medium. The boundary conditions are used to couple the system to a reservoir of
the invading phase with a constant mass flux. At the linear level of small front fluctuations, both phase field models can be
analyzed through linearized interface equations, that is, the evolution of the interface can be
described in terms of the interface profile alone. Moreover, the bulk diffusion fields implicit in this description cause the interface
equation to become spatially non-local \cite{DU99,DU00,HM01}. It is thus of interest to examine
how the models are influenced by varying disorder strength, which can be easily realized in the
experiments, too. To this end, we first define
a critical value of the disorder strength $\sigma_{c}$, above which the disorder becomes strong.
We find that at weak disorder strengths ($\sigma<\sigma_{c}$) both models have the same
interface scaling behavior given by super-rough anomalous scaling,
which is also predicted by the linear theory.
We observe clear deviation from this scaling when the disorder strength
comes close to $\sigma_{c}$, where super-roughness disappears
and weak intrinsic anomalous scaling arises. Furthermore,
in the limit of strong disorder ($\sigma>\sigma_{c}$),
the two models are no longer equivalent.
The one-sided model can still be applied to describe diffusive liquid invasion in good agreement with experimental
results of Ref.~\cite{SO02}. However, in this limit the symmetric model exhibits nucleation
of the invading phase in front of the advancing interface.
%, and thus the single-valued interface description breaks down. Instead of imbibition, interface roughening with the incorporation of disperse domains at the interface may be more relevant for solidification.

The structure of the paper is as follows. In Section ~\ref{SecII} we introduce the two
versions of the phase field model, consider the linearized interface equation (LIE)
valid in the weak disorder limit, and give an estimate for the strong disorder limit,
thus introducing a scale for the disorder strength.
Section~\ref{SecIII} defines the concepts of scaling in interface roughening,
including super-rough and intrinsic anomalous scaling.
In Setion~\ref{SecIV} we present our numerical results, and finally give our
conclusions in Section~\ref{SecV}.
%25.2.

\section{The Phase Field Model}\label{SecII}

The phase field model we are using describes a system of two
phases separated by an interface. We introduce a
locally conserved field $\phi$ to represent the two phases of the
problem with the equilibrium values $\phi_{e}=\pm 1$ in such a
way that the interface position is at
$\phi(x,h)=0$. The dynamics of the field is then assumed to follow a
conserved equation based on a time-dependent Ginzburg-Landau Hamiltonian
(model B \cite{HH77}):
\begin{equation}\label{eq:PFM_j}
\partial\phi/\partial t=-\nabla\cdot \mathbf{j},
\end{equation}
where the current is proportional to the gradient of the
chemical potential $\bold{j}=-M(\phi)\nabla\mu$. Here, the
chemical potential is $\mu=\delta \mathcal{F}/\delta\phi$ and
the free energy is taken to be of the form
$\mathcal{F}[\phi]=\int\mathnormal{d}\pmb{r}(V(\phi)+(\epsilon\nabla\phi)^{2}/2)$.
A simple double-well potential is chosen with a linear random
term:
\begin{equation}\label{eq:DoubleWell}
V(\phi)=-\frac{1}{2}\phi^{2}+\frac{1}{4}\phi^{4}-\eta(\pmb{r})\phi
\end{equation}
The variable $\eta(r)$ is taken to be stochastic and it
plays the role of spatially quenched disorder in the system. The
disorder is characterized by its average $\langle\eta\rangle$,
and its standard deviation $\sigma$, which characterizes the disorder strength.
The disorder also has a correlation length $l_{corr}$, which in a
numerical scheme is most conveniently set to the lattice spacing. Note
that by considering $\eta$ as local average over area of size $l_{corr}$,
the choice of the length $l_{corr}$ also
enters the disorder strength. That is, the standard deviation of $\eta$
is $\sigma$ when observed at scale $l_{corr}$.
The surface tension of this model can be calculated with the disorder-free
``kink'' solution $\phi_{0}$, or Goldstone mode, given by the lowest energy
solution of the boundary conditions $\phi(-\infty)=-1$ and
$\phi(+\infty)=1$. The resulting dimensionless surface tension
is $\gamma=\sqrt{2}/3\simeq 0.47$~\cite{EL01}.\\
The equation for the dynamics of the phase field reads
\begin{align}\label{eq:PFM_phi}
\frac{\partial\phi}{\partial t} & =  \bm{\nabla}
M(\phi)\bm{\nabla}\mu \nonumber \\  {} &=\bm{\nabla}
M(\phi)\bm{\nabla}\big[-\phi+\phi^{3}-\epsilon^{2}\nabla^{2}\phi-\eta(\bm{r})\big],
\end{align}
where $M(\phi)$ is a mobility parameter which may depend on the phase
field. In the \emph{sharp interface limit} $\epsilon\to 0$, the
normal velocity of the interface can be obtained by integrating
equation (\ref{eq:PFM_phi}) in a region around the interface,
\begin{equation}\label{eq:Normal velocity}
v_{n}\simeq j_{n}\vert_{\pm}=
M(\phi)\partial_{n}\mu\vert_{+}-M(\phi)\partial_{n}\mu\vert_{-},
\end{equation}
where the subscripts $+$ and $-$ correspond to the two phases of
the system. In our study, two different functional forms for the
mobility $M(\phi)$ will be considered. First, we assume a
\emph{symmetric} parameter $M=M_{0}$ constant for the whole
system independent of the field $\phi$. In this case, the velocity of the interface is
controlled by the difference between incoming and outgoing
currents. Second, we will consider a
\emph{one-sided} parameter $M=M_{0}\theta(\phi)$,
$\theta(\phi)$ being the Heaviside step function, which is zero in
the phase where $\phi<0$. Note that the normal velocity of the
interface then reduces to $v_{n}\simeq
-M_{0}\partial_{n}\mu\vert_{-}$, that is, it is only
proportional to the outgoing current. As we will see in our
numerical results, the two models can give different results and
describe different physical situations depending on the
strength of the disorder.

Here we consider the so-called driven case, where the mean velocity of the
interface is fixed to a constant value. The relevant boundary
condition is to impose a fixed gradient of the chemical
potential at the bottom of the system, $\nabla\mu=-V\hat{y}$.
Combined with the initial condition of
\begin{equation}
\label{eq:initial}
\phi=\left\{
\begin{array}{ll}
+1 &  y<H_{init};\\
-1 &  y\geq H_{init},
\end{array}
\right.
\end{equation}
the boundary condition leads to phase $\phi=+1$ invading phase
$\phi=-1$ and the interface moving with a constant average
velocity. In the general context of the phase field model we
will refer to these phases as A and B, respectively. In terms of
liquid front invasion into a Hele-Shaw cell, these phases
would be liquid and air, respectively, whereas in terms of phase
separation they would be the phases rich in components A and B.

\subsection{The linearized interface equation}\label{SubSecIIA}

The dynamics of a front in the phase field model described above can also be considered in terms of
an integro-differential equation for the interface, which reduces the dimensionality of the problem by one,
obviously a desirable property. However, this description can only be obtained as a
perturbation expansion around a flat front with small fluctuations, and since the fluctuations are caused by the
disorder, this also means weak disorder. Because the front dynamics are only obtained to first order in small
fluctuations we call it the linearized interface equation (LIE). It is noteworthy that the long-ranged
effects of mass conservation in the phase field results in the LIE being spatially non-local,
even though it is linear. This means that the LIE takes the form of convolutions in position,
which can be made local in Fourier space.

The LIE can be obtained using the Green's function $G$ to express the phase field equation,
Eq.\ (\ref{eq:PFM_phi}), in terms of an integro-differential equation \cite{DU99},
and then projecting it to an interfacial description with the operator $P[\cdot]=\int du\partial_u\phi_0(u)[\cdot]$.
The LIE takes the form of dispersion relation for Fourier
components of small front fluctuations $\hat{h}(k,t)$
around the average interface height $H_0(t)$, ({\it i.e.} $H(x,t)
= H_0(t) + h(x,t)$):
\begin{equation}
\label{eq:LIE}
\partial_t \hat{h}(k,t)=-(\dot{H}_0|k|+\gamma|k|^3)\hat{h}(k,t)
+\vert k\vert\hat{\eta}(k,t),
\end{equation}
where $\gamma$ is surface tension, $\dot{H}_0=V$ is the
(constant) average front propagation velocity, and the disorder
term is the Fourier transform of the two-dimensional (2D) disorder along the
front, or $\hat{\eta}(k,t)=\int dx e^{-ikx}\eta(x,H(x,t))$.
From the dispersion relation \eqref{eq:LIE}, one immediately
obtains the crossover length scale
\begin{equation}
\xi_{\times}=2\pi\sqrt{\frac{\gamma}{\dot{H}_0}},
\end{equation}
when the two dispersion terms are equally strong. Physically the dominant
dispersion mechanism then changes from surface tension to mass transport. In addition, because the conservative character of the capillary disorder term $\vert k\vert\hat{\eta}(k,t)$, it turns out that the crossover length $\xi_{\times}$ acts as an upper cutoff for the
correlation length of fluctuations. 
% which has been numerically shown that $\xi_{\times}$ acts as an upper cutoff for the correlation length of fluctuations. 
This means that in the long wavelength region,
where mass transport controls the dissipation of front fluctuations, the interface is asymptotically flat.
This effect has been numerically shown in Refs.~\cite{AL04,DU99,DU00,LA05}, and also in
a general context of kinetic roughening~\cite{PR07}.

\subsection{Definition of disorder strength}

In order to study the influence of disorder strength
in the phase-field model of Eq. (3), we need to define
a measure for the relative strength of the disorder.
This can be achieved by comparing
the disorder contribution in the
dimensionless bulk free energy of Eq~\eqref{eq:DoubleWell} to the
surface energy. We do this by considering a domain of linear
size $r$, where the local disorder average $\langle
\eta\rangle_{r}$ is a stochastic variable with standard deviation
$\sigma_r\propto r^{-1}\sigma$, where $\sigma$ is the standard deviation of a
single disorder site, which is of linear size $l_{corr}$. The underlying
disorder then has a correlation length $l_{corr}$.
Considering a circular domain of radius
$r$, it is energetically favorable for this domain to be of the
opposite phase than its surroundings if
\begin{equation}
\label{eq:bubble}
2\pi r\gamma \leq \pi r^2\Delta\phi\langle \eta\rangle_{r},
\end{equation}
where the miscibility gap in our model is $\Delta\phi=2$. The
LHS is the energy cost of the interface, whereas the RHS is the
energy gain due to disorder. We consider the local disorder
average $\langle \eta\rangle_{r}$ on the fluctuation site to
be as large as its standard
deviation $\sigma_r = {\sigma~l_{corr}}/{\sqrt{\pi}r}$.
Then Eq.~\eqref{eq:bubble} gives the condition
\begin{equation}
\label{eq:stronglimit}
\sigma\geq \frac{\sqrt{\pi}}{l_{corr}}\gamma,
\end{equation}
for when the disorder can locally dominate the bulk energy and
thus is defined to be strong. The order of magnitude estimate for strong
disorder in our dimensionless units ($l_{corr}=1$, $\epsilon=1$) is thus
obtained as the variance being of the same order as the surface
tension $\sigma =\sigma_{c}\approx \gamma$. Note in particular that no $r$
dependence remains in the estimate, and thus the relative
disorder strength will be the same \emph{at all length scales}
(larger than the interface width $\epsilon$ and disorder site
size $l_{corr}$).

\section{Scaling of rough interfaces}\label{SecIII}

\begin{figure*}[t!]%
\includegraphics[width=5.8cm]{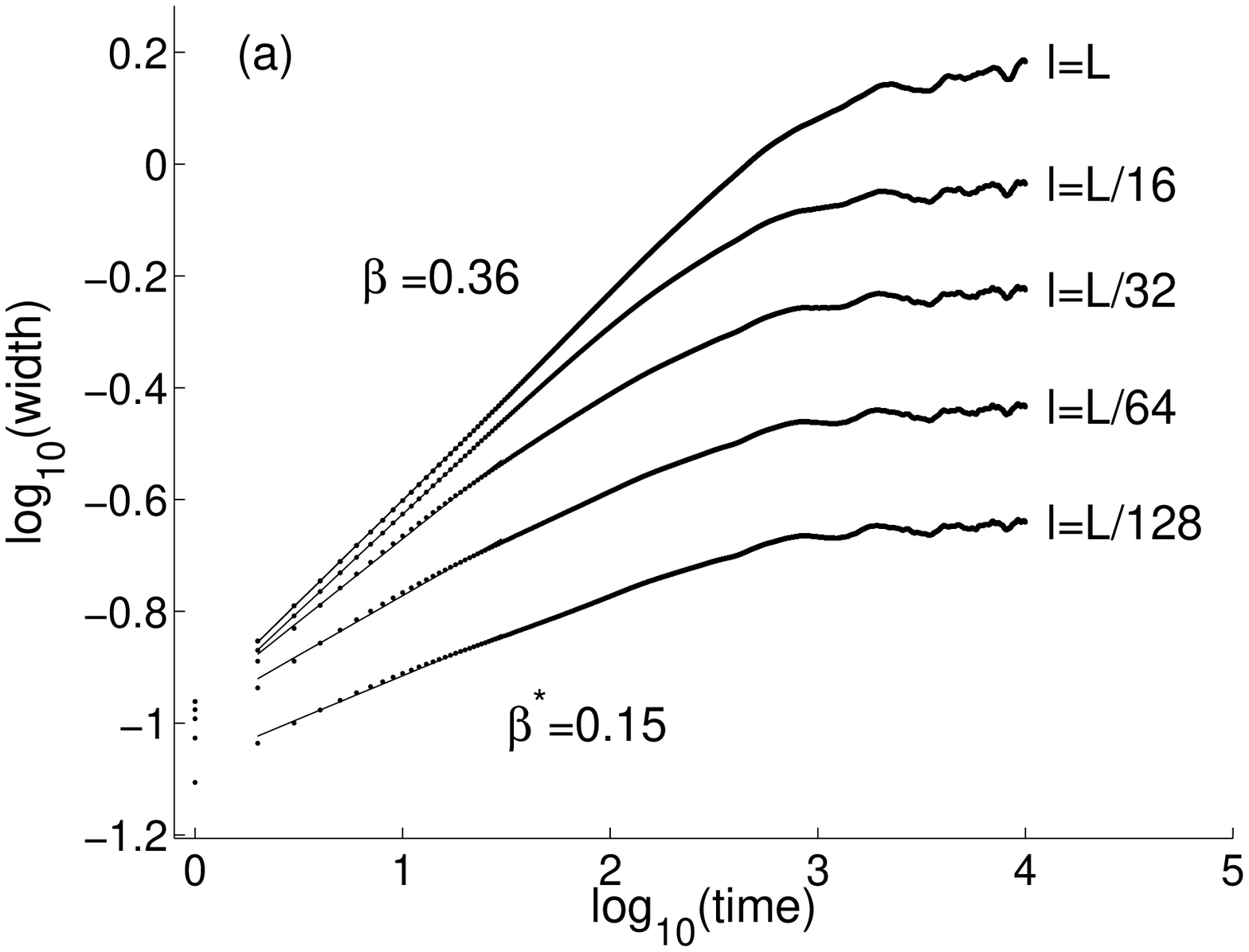}%
\includegraphics[width=5.8cm]{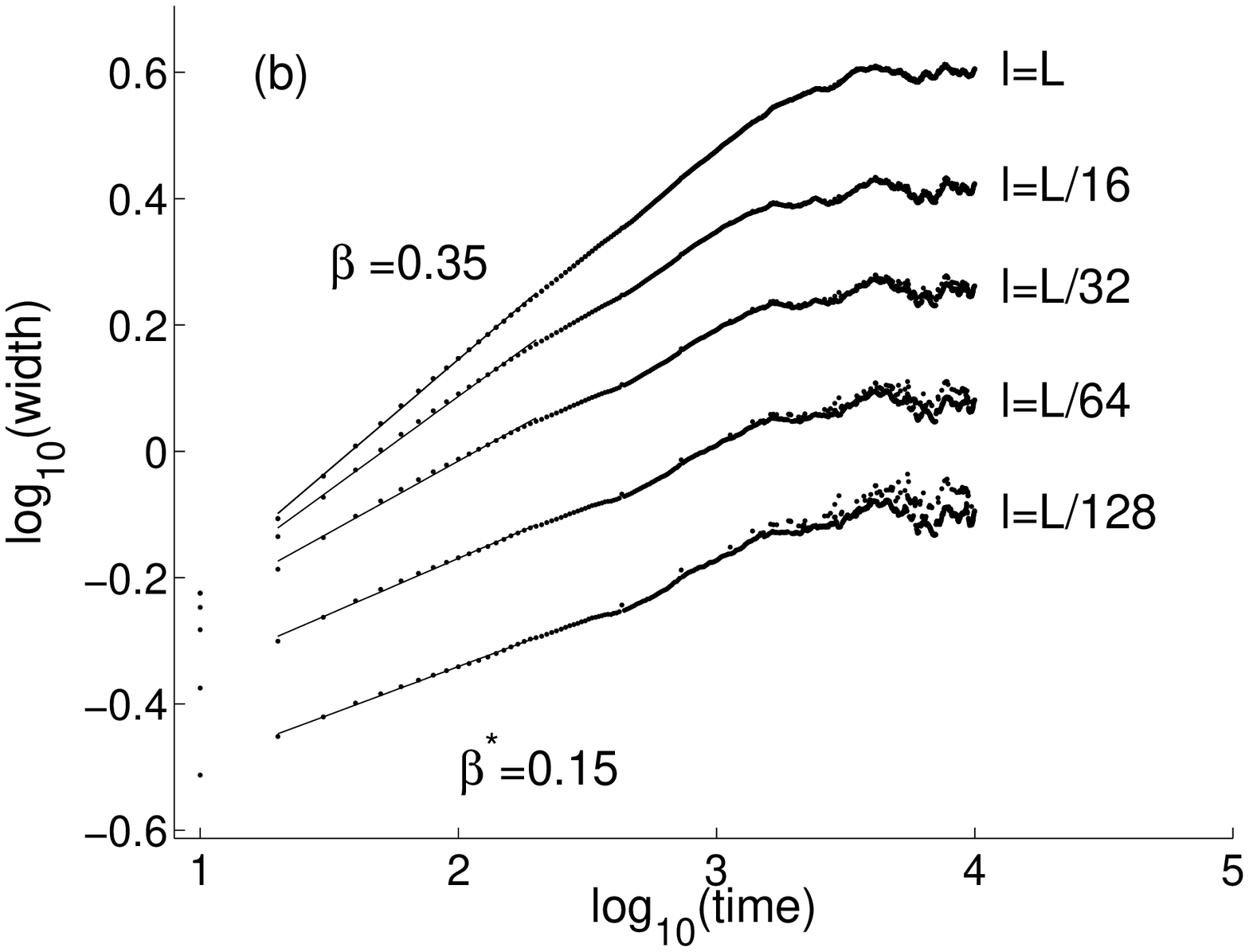}%
\includegraphics[width=5.8cm]{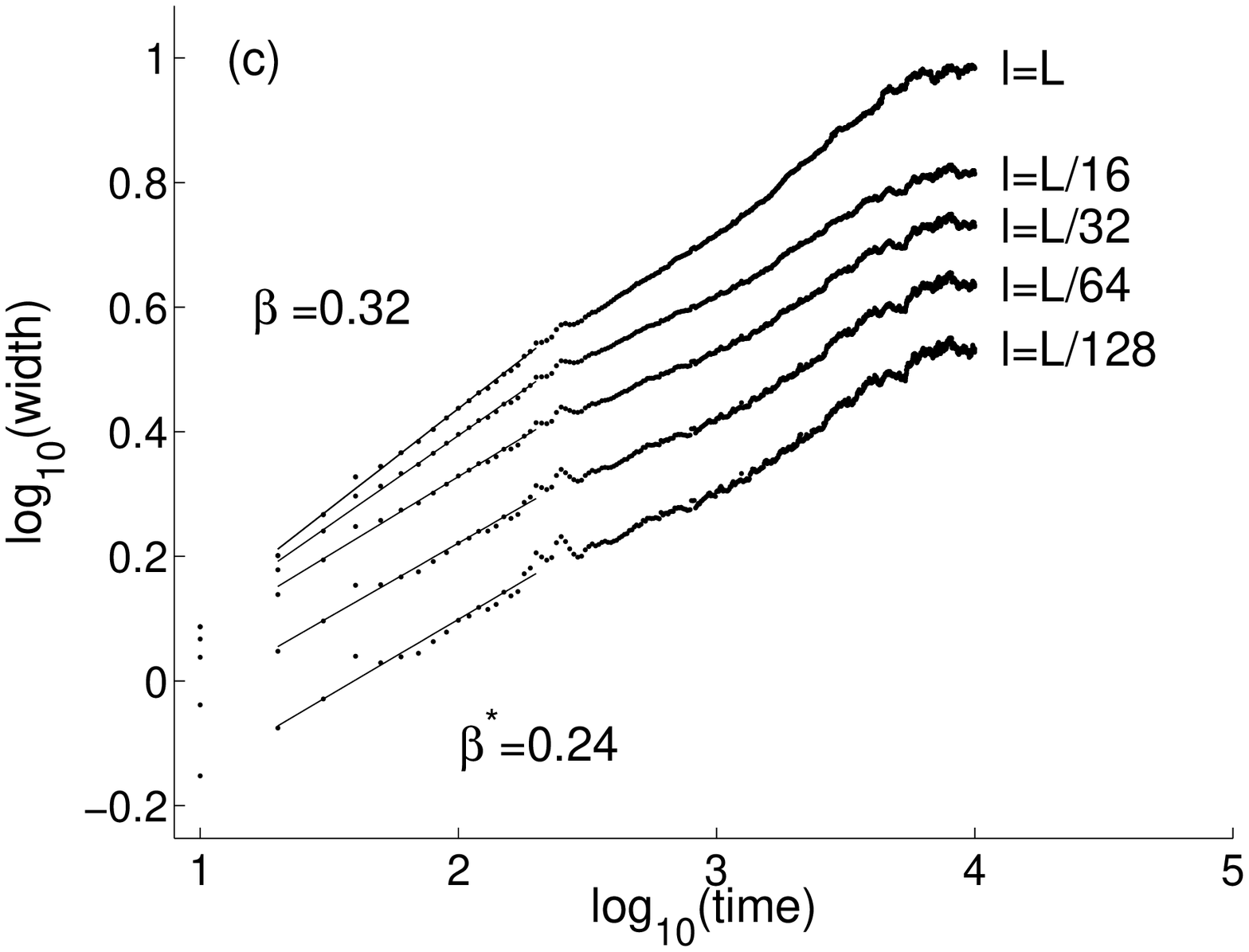}%\\
\vspace{0.2cm}
\includegraphics[width=5.7cm]{widthsWeak1SM.eps}
\includegraphics[width=5.7cm]{widthsMed1SM.eps}
\includegraphics[width=5.7cm]{widthsStrong1SM.eps}
\caption{Interface widths for the two models at different
disorder strengths (in dimensionless units). Results from the symmetric and one-sided
model are given in the upper(a,b,c) and lower(d,e,f) panels, respectively.
Disorder strength is varied from left to right as: weak
disorder ($\sigma=0.2$), intermediate disorder ($\sigma=0.5$), and strong
disorder ($\sigma=1$). Fitted growth exponents for the smallest and
largest slopes are given in the figures,
with solid lines representing the fits.}
\label{fig:widths}
\end{figure*}%

The statistical treatment of a 1D interface
$H(x,t)$ is usually done by studying the scaling properties of
its fluctuations over the whole system of total size $L$ \cite{BA95}. For
scale-invariant growth the lateral correlation length of the
fluctuations is expected to grow in time following a power law
$\ell_{c}\sim t^{1/z}$ until it reaches the system size $L$,
defining a saturation time $t_{s}\sim L^{z}$.
Alternatively, the global width of the interface $W(L,t)=\langle
\overline{[H(x,t)-\overline{H}]^{2}}\rangle^{1/2}$ increases as
$W(L,t)\sim t^{\beta}$ for $t<t_{s}$ and becomes constant
$W(L,t)\sim L^{\chi}$ for $t \geqslant t_{s}$. Here, $\langle
..\rangle$ denotes average over different noise realizations and
the overbar is an spatial average in the $x$ direction. The quantities $\chi$,
$\beta$ and $z$ are the roughness, growth and dynamical
exponent, respectively, and they are related through the scaling
relation $\chi=\beta z$. In the standard Family-Vicsek scaling
framework~\cite{FA85}, this set of scaling exponents fully
describe the scaling properties of the interface fluctuations.

However, experimental results and several growth models have
appeared in the last decade showing that global and local scales
are not always equivalent.
This is called \emph{anomalous scaling}~\cite{RA00,LO99}.
Therefore, one has to compute the interface width at different window sizes, $w(\ell,t)=\langle
\langle{[H(x,t)-\langle{H}\rangle_{\ell}]^{2}\rangle_{\ell}}\rangle^{1/2}$,
where $\langle...\rangle_{\ell}$ denotes an average over $x$ in
windows of size $\ell<L$. For scale-invariant interfaces local
fluctuations are expected to increase as
\begin{equation} \label{eq:ScalingWAnsatz}
w(\ell,t)= \ell^{\chi}g(\ell / t^{1/z}),
\end{equation}
with the corresponding scaling function
\begin{equation} \label{eq:ScalingFunctionW}
 g(u) \sim  \left\{\begin{array}{ll} u^{-(\chi-\chi_{loc})} &
     \textrm{for} \quad u<<1; \\ u^{-\chi}  &  \textrm{for} \quad u>>1,\\
       \end{array} \right.
\end{equation}
where $\chi_{loc}$ is the local rough exponent and it
characterizes the roughness at small scales. One of the
implications of anomalous scaling is that the local width
saturates when the correlation length reaches the system size,
{\it i.e.} at the time $t_{s}$ and not at the local time $t_{\ell}\sim
\ell^{z}$ as occurs in the Family-Vicsek scaling. There is an
intermediate regime between $t_{\ell}$ and $t_{s}$ where the
local width grows as $w(\ell,t)\sim t^{\beta^{*}}$ with
$\beta^{*}=\beta-\chi_{loc}/z$.

Another useful technique to
determine the whole set of scaling exponents is to study the
power spectrum of the interface
$S(k,t)=\langle\tilde{h}(k,t)\tilde{h}(-k,t)\rangle$.
In the presence of anomalous scaling it is expected
to show the following scaling
\begin{equation} \label{eq:ScalingSPCAnsatz}
S(k,t)=k^{-(2\chi+1)}s_{A}(kt^{1/z}),
\end{equation}
where the scaling function has the general form
\begin{equation} \label{eq:ScalingFunctionSPC}
 s_{A}(u) \sim  \left\{\begin{array}{ll} u^{2(\chi-\chi_{s})} &
     \textrm{for} \quad u>>1; \\ u^{2\chi+1}  &  \textrm{for} \quad
     u<<1,\\
       \end{array} \right.
\end{equation}
and $\chi_{s}$ defines the spectral roughness exponent. Different
scaling arises from this scaling function~\cite{RA00}.
For $\chi_{s}<1$ it is always valid that
$\chi_{loc}=\chi_{s}$ and the Family-Vicsek scaling is recovered
whenever $\chi_{loc}=\chi$. In contrast, the so-called intrinsic
anomalous scaling is observed when $\chi_{s}\ne\chi$. Note that
for this kind of anomalous scaling  a temporal shift in the
power spectrum is observed. The quantification of this shift
based on the above scaling form is unreliable and inaccurate,
however, and thus we refrain from giving a measure for
$\chi-\chi_s$ in the one case we observe scaling of this type.
On the other hand, for $\chi_{s}>1$
$\chi_{loc}=1$ and the super-rough
scaling appears when $\chi=\chi_{s}$.

It is worth to remember here that the crossover length $\xi_{\times}$,
present in the imbibition phenomenon becomes a very important
scale in the kinetic roughening process. More precisely,
extensive numerical studies~\cite{AL04,DU99,DU00,LA05} have shown
that fluctuations of the interface do not evolve in time beyond
this crossover length. 
%and therefore, the crossover length acts as a cutoff for the interface fluctuations. 
Therefore, the interface fluctuations reach the
steady-state at $t_{s}\simeq
\xi_{\times}^{z}$ instead of saturating at $L^{z}$. 
Additionally, since the interface is asymptotically flat at scales larger than the crossover length,
the crossover shows up as a maximum at the corresponding wavevector in the saturated
structure factor $S(k,t\rightarrow\infty)$.

% \ref{SubSecIIA}
\begin{comment}
In addition, this cutoff
manifests as a maximum in the power spectrum of
the front. This follows from the fact that at scales larger than $\xi_{\times}$ (small $k$) the average of the interface fluctuations along a lateral length $\ell$  decreases as such length $\ell$ increases, giving rise then to  a power spectrum which can be an increasing function in $k$. On the other hand, at scales smaller than $\xi_{\times}$ (large $k$), the average of the interface fluctuations increases as $\ell$ increases, and then the power spectrum follows a decreasing function in $k$ given by Eq.\  (\ref{eq:ScalingSPCAnsatz}).
%This follows from the fact that the average of uncorrelated fluctuations decrease as the length the average if taken over increases, whereas the average increases if correlations persist at the length scales considered.
\end{comment}

\section{Numerical Results}\label{SecIV}

Our study will be focused on the influence of the disorder on
the scaling behavior of the fluctuating interface for the two cases
of the mobility $M$ in the phase field model, that is, the
one-sided model and the symmetric case. In our simulations we
have used gaussian distributed disorder with zero mean
$\langle\eta\rangle=0$, and different disorder strengths
(standard deviation of $\eta$) $\sigma$. In the numerical scheme
the disorder will have a correlation length as long as the
lattice spacing, meaning that the lattice spacing normalizes the
standard deviation when dimensionless numbers in the numerical scheme
are turned to physical units.

For the driven boundary condition, the methods used to obtain the
interface description, and ultimately the LIE,
predict that the mean value of the disorder will be irrelevant,
as it has no contribution for the interface propagation.
This has been numerically verified by our
simulations of the full phase field models using different
values for the disorder average. Here we only report results
with $\langle\eta\rangle=0$.

As numerical method, we chose the simple explicit Euler scheme, where the
disorder can be straightforwardly added. In two dimensions the
timestep requirement of this method is not too restrictive for systems
large enough for present consideration.
\subsection{Weak disorder}
\begin{figure*}[t!]%
\includegraphics[width=5.8cm]{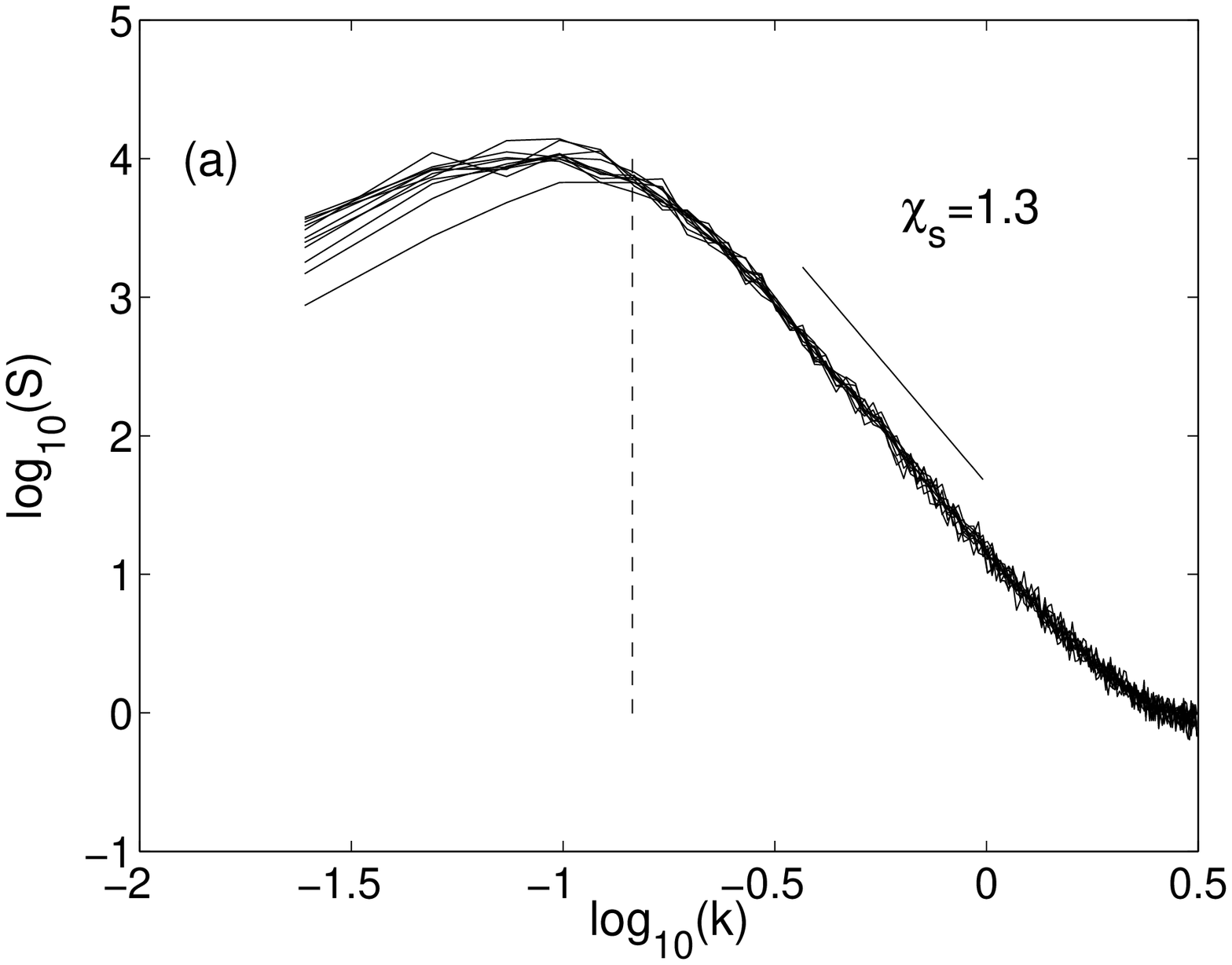}%
\includegraphics[width=5.8cm]{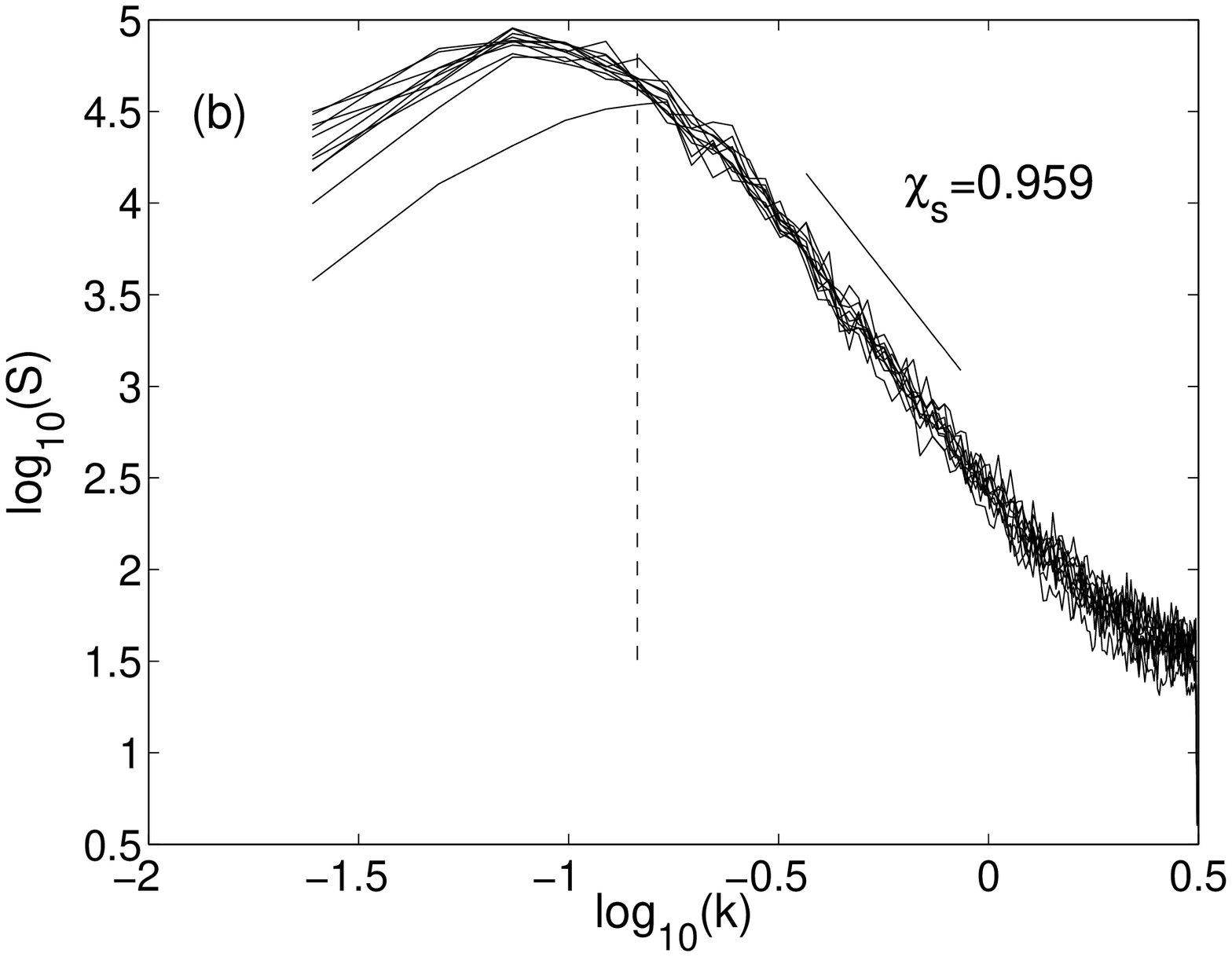}%
\includegraphics[width=5.8cm]{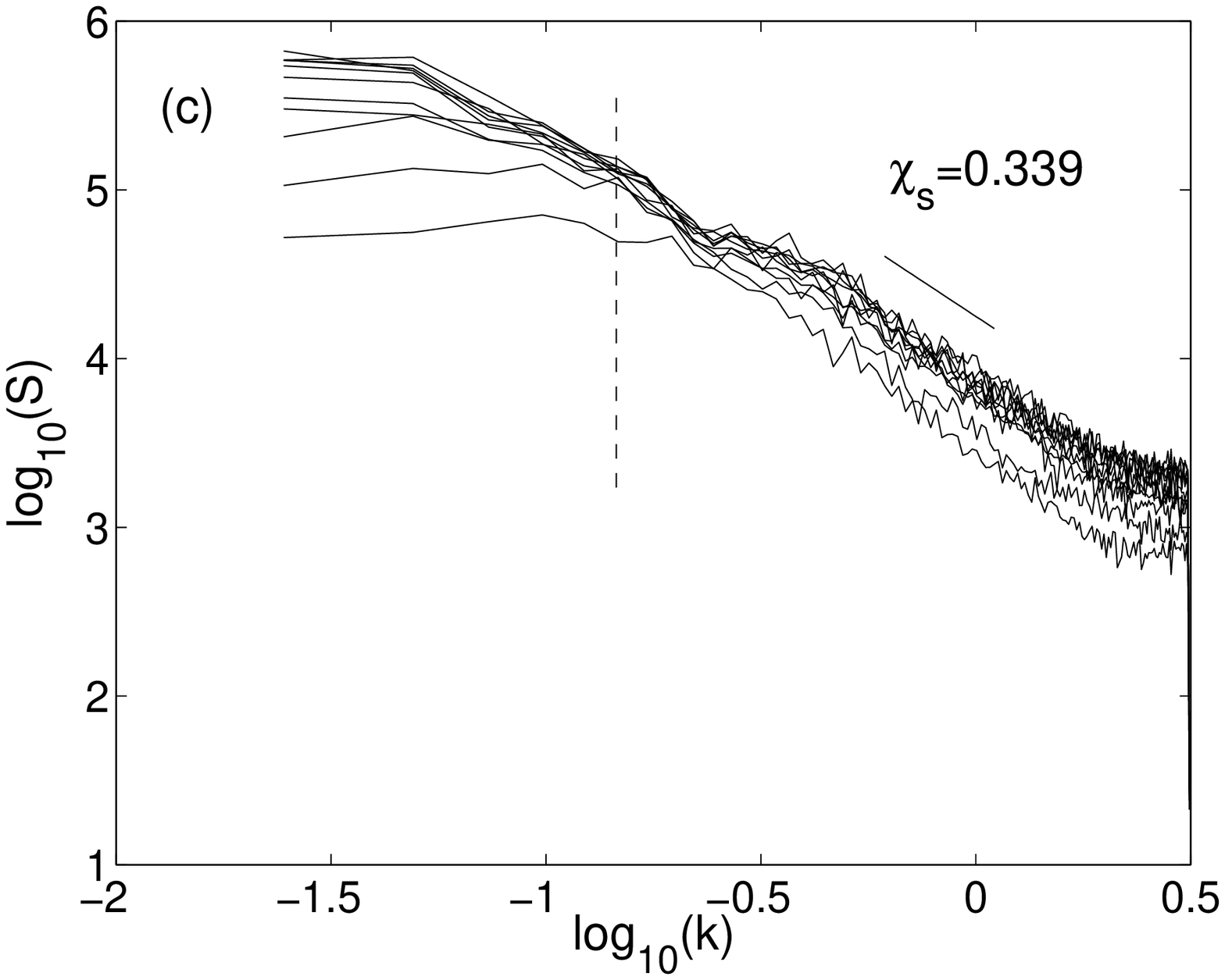}\\%
\vspace{0.3cm}
\includegraphics[width=5.6cm]{strWeak1SM.eps}%
\hspace{0.02cm}
\includegraphics[width=5.6cm]{strMed1SM.eps}%
\hspace{0.02cm}
\includegraphics[width=5.6cm]{strStrong1SM.eps}%
\caption{Structure factors at ten equidistant time intervals for
the two models at different disorder strengths (in dimensionless units). Results from the
symmetric and one-sided model are given in the upper(a,b,c) and lower(d,e,f)
panels, respectively. Disorder strengths are varied from left
to right as weak ($\sigma=0.2$), intermediate ($\sigma=0.5$) and
strong ($\sigma=1$). Fitted roughness exponents are given in the
figures, with solid lines corresponding to the fits. The dashed
vertical line corresponds to the crossover point
$k_{\times}={2\pi}/\xi_{\times}$ as
obtained from the LIE.\\}
\label{fig:strfactors}
\end{figure*}

\begin{figure}[t]%
 \centering %
 \includegraphics[width=7cm]{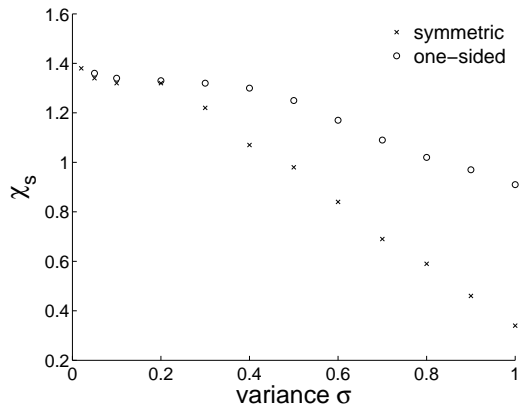}%
 \caption{Spectral roughness exponent $\chi_s$ for both phase field models as a function of
 disorder strength.}
 \label{fig:roughexp}
\end{figure}

For weak disorder, both models of the phase field
(one-sided and symmetric) are expected to be equivalent, since
the same LIE describes both cases in this limit.
The weak disorder regime 
corresponds to $\sigma\ll\gamma$, where $\gamma$ is the surface
tension. In our dimensionless units ($\epsilon=1$ in
Eq. \eqref{eq:PFM_phi}) $\gamma \simeq 0.47$. Numerically the
roughness and growth exponents of $\chi_s\simeq 1.3$ and
$\beta\simeq 0.4$, within the super-rough scaling description
($\chi_{loc}=1$, $\chi=\chi_s$), were observed by integrating the LIE~\cite{LA05}.
This is in agreement with our results for both phase field
models at weak disorder, which are shown in Figs.
\ref{fig:widths} and \ref{fig:strfactors} on the left. 
From Fig. \ref{fig:strfactors}(a) and (d)
we observe $\chi=\chi_s$, since no temporal shift in the structure
factor is present (unlike Fig. \ref{fig:strfactors}(c)).
Moreover, in Fig. \ref{fig:roughexp} we observe that at small disorder
strengths $\sigma<0.2$ in accordance with $\sigma\ll\gamma$,
the spectral roughness exponent saturates to the LIE value,
and is independent of the disorder strength.

%15.1.
We also find that numerical artifacts in the interpretation of
the interface from the phase field model appear when the
disorder is so small that the global interface width $W(L,t)$ is
much less than the numerical lattice constant. This is due to
the interpolation required to obtain the location of the
interface between two sites of the numerical grid, and shows up
as prominent periodic oscillation in interface width with
oscillation time ${\Delta x}/{\dot{H}_0}$. While this
numerical artefact can be removed by decreasing $\Delta x$, it
means that in the convenient and typically
used~\cite{DU00,HM01,AL04,LA05} dimensionless units, which lead
to Eq.~\eqref{eq:PFM_phi} with $\epsilon=1$, only a very limited
disorder strength range leads to the universal weak disorder
limit. This range is roughly $0.1<\sigma<0.2$, when
$\Delta x=1$.

%Increasing the strength of the noise away from the linear weak disorder limit we observe that the main change is in the power spectrum of the interface, see Figs. \ref{fig:roughexp} and \ref{fig:strfactors}. The roughness exponent decreases from $\chi_s \simeq 1.3$ to around $\chi_s\simeq 0.9$. It is important to remark that similar behavior in the spectral roughness exponents was also experimentally observed in Ref. \cite{SO02}, where the forced-flow imbibition process was studied in a Hele-Shaw cell.

\subsection{Strong disorder}

In the regime of strong disorder, $\sigma>0.5$, different
scenarios occur depending on which model is used. Below, we discuss the
cases of the symmetric and two-sided models separately.

\subsubsection{Symmetric model}

According to the analysis of the disorder strength above, in this
limit it is favorable for the phase field model to spontaneously
create disperse domains of one phase (A) within the region that is
initially of the other phase (B). Droplets of phase
A will form in phase B, and this mixture will initially cover
most of the system. However, local mass conservation must still
be valid regardless of any nucleation events. This means that
mass must be diffusively transported from the A phase to
the location where it nucleates within the
B phase to facilitate a growing droplet.

\begin{figure}[t]%
 \centering %
 \includegraphics[width=0.38\textwidth,origin=c,angle=0]{Int1SMS1.eps}\\%
 \includegraphics[width=0.38\textwidth,origin=c,angle=0]{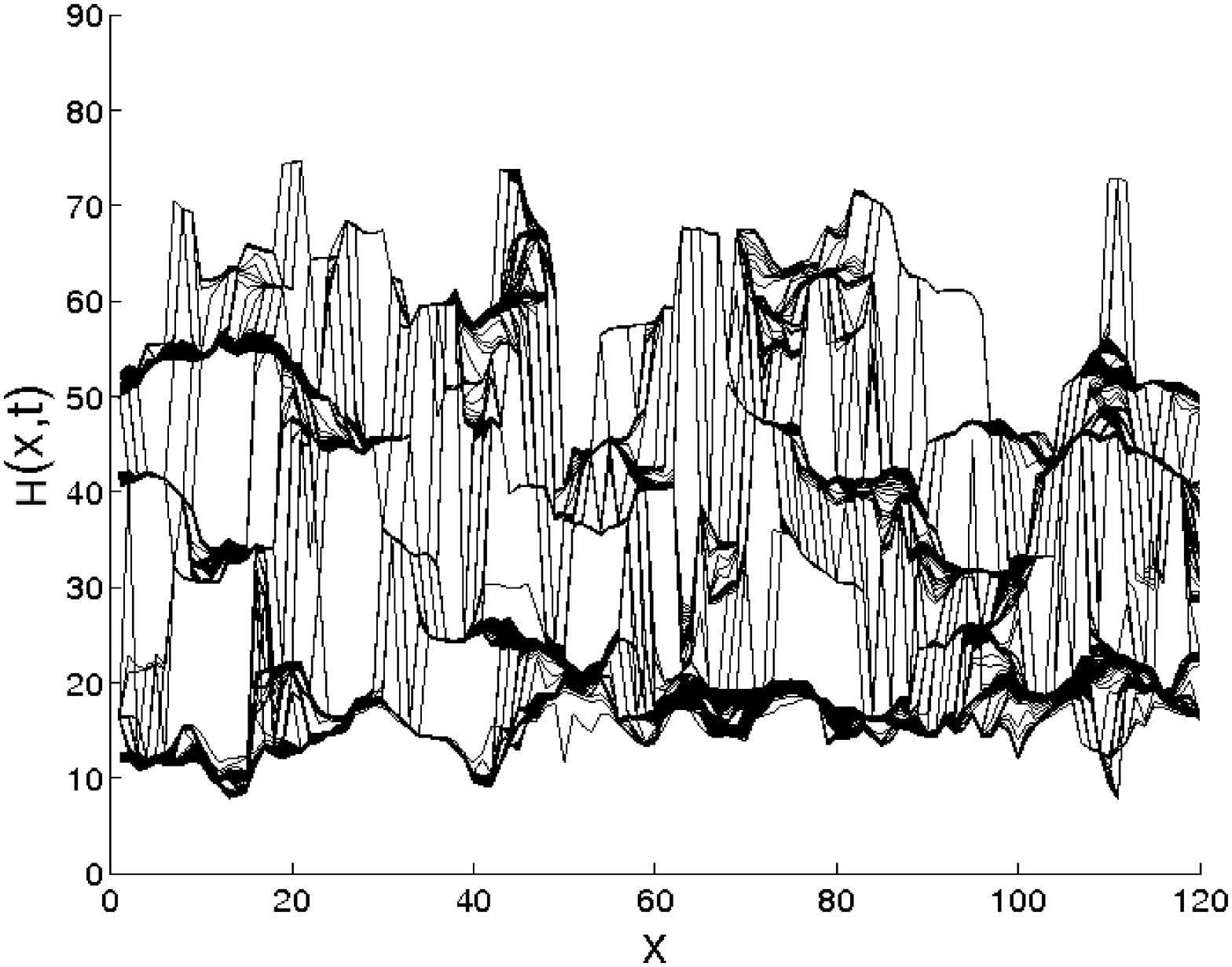}
 \includegraphics[width=0.33\textwidth, height=0.17\textwidth]{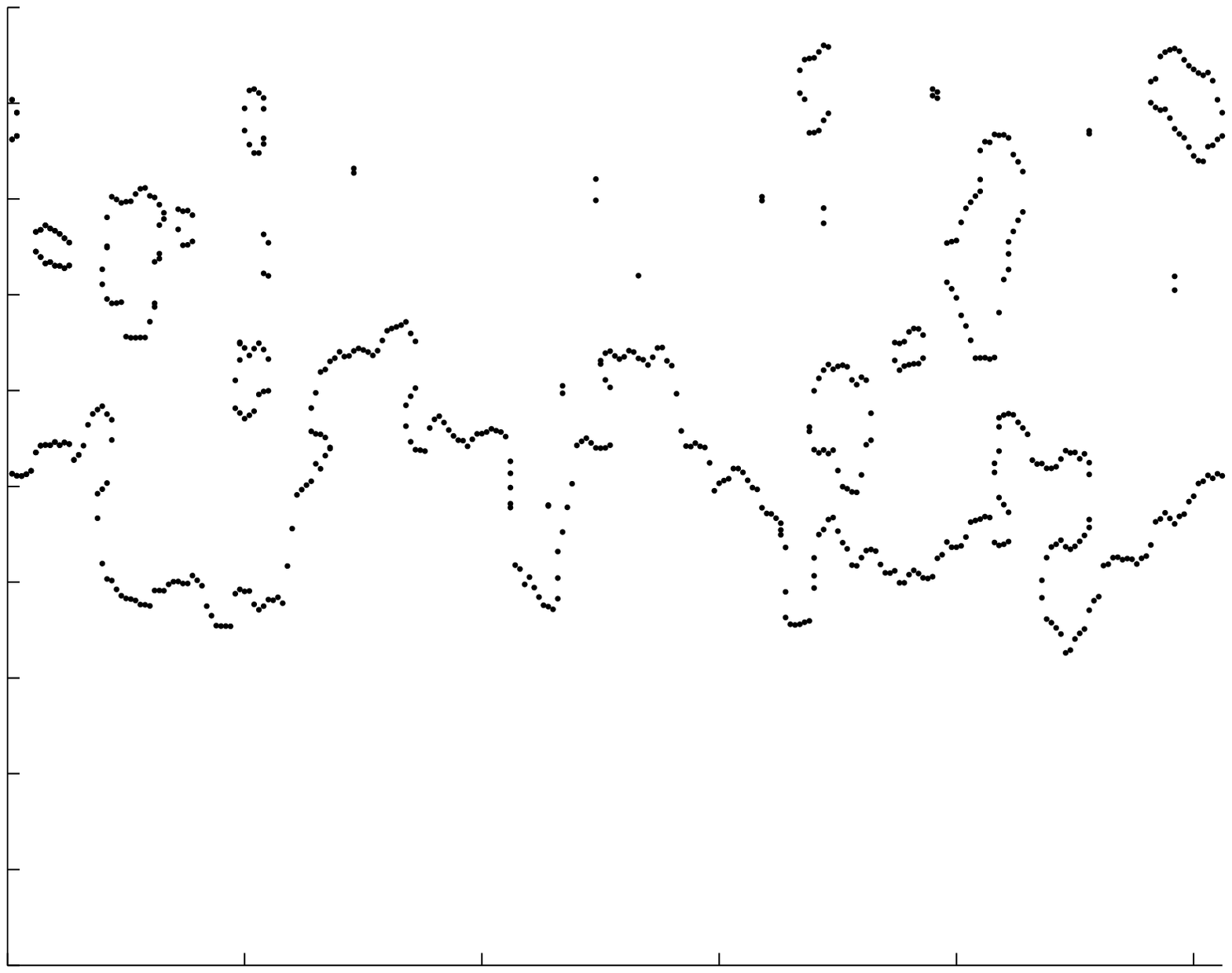}
 \caption{An example of a set of rough fronts for strong disorder
   $\sigma=1$ using both models: one-sided model (top) and symmetric model (middle).
   The bottom figure shows points of zero $\phi$ at fixed time in the symmetric model,
   demonstrating the nucleating domains.}\label{fig:INT_1SMS1}
\end{figure}

This is exactly what happens in the symmetric model (see Fig.\ \ref{fig:INT_1SMS1}), when one imposes
the initial condition of Eq. \eqref{eq:initial}. Since there is no
characteristic scale for the domain creation, the droplets are not
restricted by the crossover length $\xi_{\times}$, which acts as a
\emph{cutoff} for the interface fluctuations and therefore, the
surface roughening at large scales is different respect to a weaker
disorder strength. This is observed in Fig.\ \ref{fig:strfactors}(c),
where the interface power spectrum is plotted for a higher disorder strength at
different times using the symmetric model. We can see that the
fluctuations are not saturated, indicating that the crossover length
$\xi_{\times}$ (represented by the dashed line) is not acting anymore
as an upper cutoff. In addition, our numerical results for the
symmetric model at strong disorder show three differences to the weak
disorder case. First, the local growth exponent $\beta^{*}$ approaches
the global exponent $\beta$ (see Fig.\ \ref{fig:widths}(c)). Second, the
spectral roughness exponent decreases drastically to the range of
$\chi_s \simeq 0.5$ (see Fig.\ \ref{fig:roughexp}) and third, a temporal shift appears in the power
spectrum (see Fig.\ \ref{fig:strfactors}(c)). We can thus conclude that the scaling picture of interface
fluctuations changes from superrough to intrinsic anomalous scaling,
where $\chi_s\neq \chi$.

%These droplets change the character of surface roughening at large length scales. This is because they have no characteristic length scale, and thus are not restricted by the crossover length $\xi_{\times}$.
%
%At this point, one should note that in terms of propagation of a front of liquid into air, such as the Hele-Shaw experiment, the nucleation events are entirely unphysical. Nucleation of the advancing phase is possible in the case of solidification with impurities in the liquid phase, which induce heterogeneous nucleation. However, this case is quantitatively different from our model, because in solidification the droplets will have characteristic size (which is time dependent). Thus we don't expect that the symmetric model would quantitatively describe solidification at strong disorder.
%
%Our numerical results for the symmetric model at strong disorder show four differences to the weak disorder case. First, the local growth exponent $\beta^{*}$ approaches the global exponent $\beta$. Second, the spectral roughness exponent decreases drastically to the range of $\chi \approx 0.5$. Third, a temporal shift appears in the power spectrum. Finally, the length scale $\xi_{\times}$ no longer serves as an upper cutoff for the fluctuation correlation length. The last fact can be explained by the lack of characteristic length scale in the disperse domain sizes. The other three can be explained by the changing of the scaling picture from super-rough to intrinsic anomalous scaling.
%
The picture becomes problematic for strong disorder, however,
because the interface becomes less and less representable by a
single valued function $H(x,t)$. This is due to the interfacial
area becoming more and more tattered by overhangs, droplets and
bubbles. This also means that some numerical tricks are needed
to distinguish the interface from these bubbles and droplets.
This distinction is essentially made by finding a path for the
phase boundary across the system that locally has as small height
jumps as possible. This works relatively well as long as the
disorder is not much stronger than $\sigma=1$ in our dimensionless
units. However, note anomalous fluctuations in Fig.~\ref{fig:widths}(c)
around value $2.6$ on the vertical axis, that are due to the
abovementioned reason.
% 21.1.
% 25.2.

\subsubsection{One-sided model}

Using the one-sided model allows us to suppress the domain creation
in phase B, where the mobility parameter is zero. Then, the position of the interface $H(x,t)$ can be found by taking the largest height where the phase A has advanced to at time $t$, coming from the phase B when the phase field is above zero.
%The interface profile defined in this way cuts off any possible overhangs, and thus will give a bad description of the actual interface when the disorder is very strong. Because of the suppression of  nucleation, the one-sided model is still expected to correspond to liquid front dynamics at strong disorder. 
In Fig.\ \ref{fig:INT_1SMS1} we show 
an example of the interface profile at different times for a strong disorder, $\sigma=1.0$.

The growth exponent $\beta \simeq 0.5$ measured for strong disorder 
strength (see Fig.\ \ref{fig:widths}(e) and (f)) agrees with the experimental 
value of $\beta=0.50\pm 0.02$ reported in Ref.\ \cite{SO02} for liquid front dynamics into a Hele-Shaw cell. Likewise, a similar variation in the spectral roughness exponent $\chi_{s}$, which changes from $\chi_{s}\simeq 1.23$ to $\chi_{s}\simeq 0.91$ when the disorder strength is increased (see Fig.\ \ref{fig:roughexp}),  
was also experimentally observed in the same reference, whith a variation from $\chi_{s}= 1.1\pm 0.1$ to $\chi_{s}= 0.9\pm 0.1$ when the capillary forces of the Hele-Shaw cell were increased (see Fig.\ 15 in Ref.\ \cite{SO02}). On the other hand, we numerically observe that the crossover length $\xi_{\times}$ still acts as a 
cutoff length for the interface fluctuations at strong disorder (see Fig.\ \ref{fig:strfactors}).
These results indicate 
that the model can still describe the imbibition phenomenon at strong disorder.
%25.2.
%

%
\section{Conclusions and Discussion}\label{SecV}

In this work, we have studied two different ways of considering
the influence of the mobility parameter in a
Model B type of phase field model with a Ginzburg-Landau type free
energy. The main experimental context considered here is liquid front
invasion into a Hele-Shaw cell with quenched disorder \cite{SO02}. We have
focused on the case of driven front invasion, where there is a
forced constant mass flux into the system that follows locally conserved dynamics.
%This type of experiment has been studied in Ref.\ \cite{SO02}.
The symmetric model, studied for example in Refs.\ \cite{DU00,AL04,LA05}, uses a
constant mobility factor, whereas the one-sided model, studied
for example in Refs. \cite{HM01,pradas06}, uses a mobility that is
zero in the receding phase, which we call phase B. 

We note that both models have previously been turned into
non-local linear interface equations (LIEs) in the limit of
small front fluctuations, which is equivalent to weak disorder
\cite{DU00,HM01,LA05}. These LIEs
are identical for both models, and therefore
both models are expected to have identical scaling behavior
at the weak disorder limit. 
This is verified by direct comparison of numerical simulations. 
Furthermore, these results also agree with the relevant Hele-Shaw
experiments~\cite{SO02}.

We give an estimate for the strong disorder limit by comparing
the disorder contribution to bulk energy to the surface tension.
We find that the linear weak disorder limit is found well below
this disorder value, and that only in this limit does the roughness
exponent not continuously depend on the disorder strength
in either model. This means that a well-defined region of universality
only exists at the weak disorder limit.

Numerically we study the dependence of the growth and roughness exponents
of the invasion fronts as a function of the disorder strength. Our results
are consistent with a (continuous) change of scaling behavior from superrough to intrinsic
anomalous scaling, when the disorder strength is increased from weak to strong.

At strong disorder,
the symmetric model is no longer found to correspond to the
Hele-Shaw experiment due to domain creation of the invading phase in front
of the propagating interface. As our analysis shows, the domain growth can occur at all length scales (larger than the disorder site size $l_{corr}$ of the system) without any characteristic radius, which is a phenomenon observed in other experimental situations such as nucleation on dislocations~\cite{cahn57} or binary mixtures~\cite{GU83}. 
%This nucleation qualitatively resembles heterogeneous nucleation in driven solidification, but our model doesn't include the relevant properties of heterogeneously nucleated domains.
In contrast, the results for the one-sided model do agree well with the Hele-Shaw
experiments~\cite{SO02}, even as the disorder strength is increased.

We hypothesize that the change in scaling behavior is due to decreased
effect of surface tension and mass transport in interface roughening as the
disorder becomes strong. Since the argument of strong disorder is the
dominance of disorder over surface tension, the scaling in the regime dominated by the surface tension (characterized by having a correlation length $\ell_{c}<\xi_{\times})$ should change for both models at strong disorder.
In the one-sided model, mass transport still restricts interface roughening, and thus
the crossover scale $\xi_{\times}$ persists. Conversely in the symmetric model the
nucleated domains create roughening by avalanches that are not controlled by
mass transport from the reservoir, and thus the crossover scale $\xi_\times$
becomes irrelevant.
%11.3.
%
\begin{acknowledgments}
This work was supported by the DGI of the Ministerio de Educaci{\'o}n y Ciencia
(Spain) through Grant Nos. FIS2006-12253-C06-05, and by the Academy of Finland through
the TransPoly consortium and COMP Center of Excellence grants. Some of the computations
presented in this document have
been made with CSC's computing environment. CSC is the Finnish IT center
for science and is owned by the Ministry of Education. 
\end{acknowledgments}


\begin{thebibliography}{99}
\bibitem{cahn65}{J.W. Cahn \emph{J. Chem. Phys.} \pmb{42}, 93 (1965).}

\bibitem{GU83}{J.D. Gunton, M. San Miguel, and P.S. Sahni, in \emph{Phase Transitions and Critical Phenomena}, edited by C. Domb and J.L. Lebowitz (Academic, New York, 1983), Vol. 8, p. 267.}

\bibitem{langer80} J.S. Langer \emph{Rev. Mod. Phys.} \pmb{52}, 1,(1980).

\bibitem{boettinger02}W. J. Boettinger, J. A. Warren, C. Beckermannand and A. Karma, \emph{Annu. Rev. Mater. Res.} \pmb{32}, 163 (2002).

\bibitem{AL04}{M. Alava, M. Dub\'e and M. Rost, \emph{Adv. Phys.} \pmb{53}, 83 (2004).}

\bibitem{AN04}{T. Ala-Nissila, S. Majaniemi and K. Elder, \emph{Lect. Notes Phys.} \pmb{640}, 357 (2004).}

\bibitem{HH77}{P.C. Hohenberg and B.I. Halperin, \emph{Rev. Mod. Phys.} \pmb{49}, 435 (1977).}

\bibitem{elder94} K.R. Elder, F. Drolet, J. M. Kosterlitz, M. Grant, \emph{Phys. Rev. Lett.} \pmb{72}, 677 (1994).

\bibitem{EL01}{K.R. Elder, M. Grant, N. Provatas and J.M. Kosterlitz, \emph{Phys. Rev. E} \pmb{64}, 021604 (2001).}

\bibitem{chen02}{L. Q. Chen, \emph{Annu. Rev. Mater. Res.} \pmb{32}, 113 (2002).}

\bibitem{LU05}{K. Luo, M.-P Kuittu, C. Tong, S. Majaniemi, and T.Ala-Nissila, \emph{J. Chem. Phys.} \pmb{123}, 194702 (2005).}

\bibitem{LA06}{T. Laurila, C. Tong, S. Majaniemi, and T. Ala-Nissila, \emph{Phys. Rev. E }\pmb{74}, 041601 (2006).}

\bibitem{rost07}M. Rost, L. Laurson, M. Dub\'e and M. Alava, \emph{Phys. Rev. Lett.} \pmb{98}, 054502 (2007).

\bibitem{SO07}{R. Planet, M. Pradas, A. Hern\' andez-Machado and J. Ort\'in, \emph{Phys. Rev. E} \pmb{76}, 056312 (2007).}

\bibitem{SO05}{J. Soriano, A. Mercier, R. Planet, A. Hern\'andez-Machado, M.A. Rodr\'iguez and J. Ort\'in, \emph{Phys. Rev. Lett.} \pmb{95}, 104501 (2005).}

\bibitem{SO02}{J. Soriano, J. Ort\'in and A. Hern\' andez-Machado, \emph{Phys. Rev. E} \pmb{66}, 031603 (2002).}

\bibitem{SO02b}{J. Soriano, J.J. Ramasco, M.A. Rodr\'iguez, A. Hern\'andez-Machado and J. Ort\'in, \emph{Phys. Rev. Lett.} \pmb{89}, 026102 (2002).}

\bibitem{GE02}{D. Geromichalos, F. Mugele and S. Herminghaus, \emph{Phys. Rev. Lett.} \pmb{89}, 104503 (2002).}

\bibitem{HM01}{A. Hern\'andez-Machado, J. Soriano, A.M. Lacasta, M.A. Rodr\'iguez, L. Ram\'irez-Piscina and J. Ort\'in,
\emph{Europhys. Lett.} \pmb{55}, 194 (2001).}

\bibitem{DU99}{M. Dub\'e, M. Rost, K.R. Elder, M. Alava, S. Majaniemi, and T. Ala-Nissila, \emph{Phys. Rev. Lett.} \pmb{83}, 1628 (1999).}

\bibitem{DU00}{M. Dub\'e, M. Rost and M. Alava, \emph{Eur. Phys. J. B} \pmb{15}, 691 (2000); M. Dub\'e, M. Rost, K.R. Elder, M. Alava, S. Majaniemi, and T. Ala-Nisssila, \emph{ibid.} \pmb{15}, 701 (2000).}

\bibitem{LA05}{T. Laurila, C. Tong, I. Huopaniemi, S. Majaniemi, and T. Ala-Nissila, \emph{Eur. Phys. J. B} \pmb{46}, 553 (2005).}

\bibitem{PR07}{M. Pradas, J.M. L\'opez, and, A. Hern\' andez-Machado, \emph{Phys. Rev. E} \pmb{76}, 010102(R) (2007).}

\bibitem{BA95}{A.-L. Barab\' asi, H.E. Stanley, \emph{Fractal Concepts in Surface Growth} (Cambridge University Press, Cambridge, England, 1995).}

\bibitem{FA85}{F. Family and T. Vicsek, \emph{J. Phys. A} \pmb{18}, L75 (1985).}

\bibitem{RA00}{J.J. Ramasco, J.M. L\'opez and M.A. Rodr\'iguez, \emph{Phys. Rev. Lett} \pmb{84}, 2199 (2000).}

\bibitem{LO99}{J.M. L\'opez, \emph{Phys. Rev. Lett.} \pmb{83}, 4594 (1999).}

\bibitem{pradas06}M. Pradas and A. Hern\' andez-Machado, \emph{Phys. Rev. E} \pmb{74}, 041608 (2006).

\bibitem{cahn57}{J. W. Cahn \emph{Acta Metall.} \pmb{5}, 169 (1957).}

\end{thebibliography}
\end{document}